\documentclass[12pt]{aastex}
\usepackage{natbib}
\usepackage{lscape}
\begin{document}
\title{LHS~2803B: A very wide mid-T dwarf companion to an old~M~dwarf identified from Pan-STARRS1}
\shorttitle{}
\author{Niall R. Deacon,\altaffilmark{1,2 }\email{deacon@mpia.de}
Michael C. Liu,\altaffilmark{2,3}
Eugene A. Magnier,\altaffilmark{2}
Brendan P. Bowler,\altaffilmark{2}
Andrew W. Mann,\altaffilmark{2}
Joshua A. Redstone,\altaffilmark{4}
William S. Burgett,\altaffilmark{2}
Ken C. Chambers,\altaffilmark{2}
Klaus W. Hodapp,\altaffilmark{5}
Nick Kaiser,\altaffilmark{2}
Rolf-Peter Kudritzki,\altaffilmark{2}
Jeff S. Morgan,\altaffilmark{2}
Paul A. Price,\altaffilmark{6}
John L. Tonry,\altaffilmark{2}
Richard J. Wainscoat\altaffilmark{2}}
\altaffiltext{1}{Max Planck Institute for Astronomy, Koenigstuhl 17, D-69117 Heidelberg, Germany}
\altaffiltext{2}{Institute for Astronomy, University of Hawai`i, 2680 Woodlawn Drive, Honolulu, HI 96822, USA}
\altaffiltext{3}{Visiting Astronomer at the Infrared Telescope Facility, which is operated by the University of Hawaii under Cooperative Agreement no. NNX-08AE38A with the National Aeronautics and Space Administration, Science Mission Directorate, Planetary Astronomy Program}
\altaffiltext{4}{Facebook, 1601 Willow Road, Menlo Park, CA 94025, USA}
\altaffiltext{5}{Institute for Astronomy, University of Hawai`i, 640 N. Aohoku Place, Hilo, HI 96720, USA}
\altaffiltext{6}{Princeton University Observatory, 4 Ivy Lane, Peyton Hall, Princeton University, Princeton, NJ 08544, USA}
 \label{firstpage}
 \begin{abstract}
We report the discovery of a wide ($\sim$1400~AU projected separation), common proper motion companion to the nearby M dwarf LHS~2803 (PSO~J207.0300$-$13.7422). This object was discovered during our census of the local T dwarf population using Pan-STARRS1 and 2MASS data. Using IRTF/SpeX near-infrared spectroscopy, we classify the secondary to be spectral type T5.5. University of Hawai`i 2.2m/SNIFS optical spectroscopy indicates that the primary  has a spectral type of M4.5, with approximately solar metallicity and no measurable H$\alpha$ emission. We use this lack of activity to set a lower age limit for the system of 3.5 Gyr. Using a comparison with chance alignments of brown dwarfs and nearby stars, we conclude that the two objects are unlikely to be a chance association. The primary's photometric distance of 21 pc and its proper motion implies thin disk kinematics. Based on these kinematics and its metallicity, we set an upper age limit for the system of 10 Gyr. Evolutionary model calculations suggest that the 
secondary has a mass of 72$\pm^{4}_{7}$ M$_{Jup}$, temperature of 1120$\pm$80 K, and 
$\log g$=5.4$\pm$0.1 dex. Model atmosphere fitting to the near-IR spectrum 
gives similar physical parameters of 1100 K and $\log g$=5.0.
\end{abstract}
 \section{Introduction}
The past two decades have seen an explosion of interest in the study of ultracool dwarfs (spectral types of M6 or later), which can be stars or brown dwarfs. However, determining the physical properties of brown dwarfs, such as their mass, age and radius, is complicated due to their substellar nature. Unlike main-sequence stars, the lack of hydrogen burning means that brown dwarfs have no stable luminosity and hence there is a degeneracy between effective temperature, mass and age. This degeneracy can be broken for two classes of objects. The first type, \citep[``mass benchmarks'';][]{Liu2008}, are binary brown dwarfs whose dynamical mass can be measured from their orbits. There are over a dozen such systems known \citep{Dupuy2011}. 

The second type, ``age benchmarks'', have their ages inferred from associated stars. This can be done by studying low-mass members of stellar clusters. However, the confirmed, young substellar populations of such associations are currently limited to mid-M to early-L dwarfs \citep[e.g. ][]{Lodieu2012} and is hence not directly applicable to cooler field objects. Another source of benchmarks is substellar companions to higher mass stars. These objects have their ages determined from their stellar primaries and span a wide range in spectral types \citep{Faherty2010}. Three close T dwarf companions have been discovered by high contrast imaging --- Gl~229B \citep{Nakajima1995},
GJ~758B \citep{Thalmann2010}, and SCR~1845$-$6357B \citep{Biller2006} --- along with eleven wide companions to main sequence stars discovered using wide-field surveys or conventional imaging  --- GL~570D
\citep{Burgasser2000}, HD~3651B \citep{Mugrauer2006,Luhman2007},
HN~Peg~B \citep{Luhman2007}, Wolf~940B \citep{Burningham2009}, Ross~458C
\citep{Goldman2010}, HIP~63510C \citep{Scholz2010B}, HIP~73786B
\citep{Scholz2010B,Murray2011}, $\epsilon$~Indi~Bab
\citep{Scholz2003, McCaughrean2004}, Gl~337CD \citep{Wilson2001,Burgasser2005}, HIP~38939B \citep{Deacon2012}, and BD~+01$^{\circ}$~2920B \citep{Pinfield2012}. Additionally \cite{Albert2011} identify a potential companion to HD~15220 (CFBDS~J022644$-$062522) while \cite{Dupuy2012} establish ULAS~1315$+$0826~\citep{Pinfield2008} as a potential companion to TYC 884-383-1. Finally, two ultracool companions have been discovered to white dwarfs, LSPM 1459+0857B \citep{Day-Jones2010} and WD~0806$-$661b \citep{Luhman2011}, the latter of which has no spectroscopic data but is likely to be of the newly discovered Y spectral class~\citep{Cushing2011}. These benchmarks can be used to test atmospheric and evolutionary models for mutual consistency by comparing their expected effective temperatures from their age and luminosity to the effective temperatures from atmospheric model fits to their spectra. The analysis of benchmark T dwarfs presented in \cite{Deacon2012} indicates that the most recent atmospheric models \citep{Allard2010,Saumon2008} match the effective temperatures inferred from evolutionary models to within $\sim$100 K.

The vast majority of wide brown dwarf companions have been discovered by mining large astronomical surveys. One of the leading survey telescopes currently in operation is Pan-STARRS1 \citep[PS1, ][]{Kaiser2002}. This 1.8m wide-field telescope sits atop Haleakal\={a} on Maui in the Hawaiian Islands. The facility is operated by a consortium of US, German, UK and Taiwanese institutions and has been conducting full survey operations since May 2010. The telescope carries out a variety of astronomical surveys covering a wide range of science areas, from the solar system \citep{Hsieh2012}, to the solar neighborhood \citep{Deacon2011,Liu2011}, to high redshift supernovae \citep[e.g. ][]{Chomiuk2012}. The most valuable of these surveys for ultracool dwarf science is the 3$\pi$ Survey. This is covering the three-quarters of the sky ($\delta > -30^{\circ}$) visible from Maui in five filters in the Pan-STARRS1 system \citep[$g_{P1}$, $r_{P1}$, $i_{P1}$, $z_{P1}$, $y_{P1}$, ][]{Tonry2012}. Two pairs of images will be taken per year in each passband for each area of sky during each of the three years of planned survey operation. Due to the fact that Pan-STARRS1 is only now finishing its second year of survey operation, our search for ultracool T dwarfs using PS1 has so far used data from the 2MASS survey \citep{Skrutskie2006} as an additional epoch for proper motion determination and as a source of near-infrared photometry. This has resulted in the discovery of four relatively bright ($J \sim 16.5$ mag) T dwarfs in PS1 commissioning data \citep{Deacon2011} and, in combination with data from the Widefield Infrared Survey Explorer \citep[WISE. ][]{Wright2010}, the identification of PSO~J043.5395+02.3995, a  nearby T8 brown dwarf with an extremely high proper motion (\citealt{Liu2011} see also \citealt{Scholz2011,Kirkpatrick2011}). Using the proper motion catalog produced for the PS1-2MASS T dwarf search, a dedicated search for cool companions to main sequence stars is also underway. This has already resulted in the discovery of HIP~38939B, a T4.5 companion to a mid-K star in the solar neighborhood \citep{Deacon2012}.

Here we present the discovery of a mid-T-type companion\footnote{This object was independently discovered by Muzic et al. AJ submitted\nocite{Muzic2012}.} to the M dwarf LHS~2803 (PSO~J207.0300$-$13.7422). The primary is of a much lower mass than most other stars with a wide ($>$1000 AU) T dwarf companion. We use spectroscopic observations to characterize the both the primary and secondary stars. These data are then used to fit substellar evolutionary and atmospheric models to the secondary, allowing a comparison of the best fit models.

\section{Identification in Pan-STARRS1 data}
For the past two years, we have undertaken a survey for ultracool, methane-bearing T~dwarfs in the Solar Neighborhood (\citealt{Deacon2011}, Liu et al. in prep.) combining Pan-STARRS1, 2MASS \citep{Skrutskie2006} and UKIRT/UKIDSS \citep{Lawrence2007} data. The selection process used is similar to that outlined in \cite{Deacon2011}. Objects with apparent proper motion between Pan-STARRS1 and 2MASS, red $y_{P1}-J_{2MASS}$ and $z_{P1}-y_{P1}$ colors, and a blue $J_{2MASS}-H_{2MASS}$ color are selected as the initial sample. These candidates had proper motions calculated from their 2MASS and PS1 astrometry assuming a systematic error floor between the two surveys of 80 milliarcseconds. After anti-matching with optical plate data \citep{Monet2003,Hambly2001}, the candidates are then visually inspected to remove spurious associations. Objects without UKIDSS data are then targeted for near-infrared photometry. We then obtain spectroscopy for objects selected from this process.

\begin{figure}[htbp]
\begin{center}
\epsscale{1.0}
\plotone{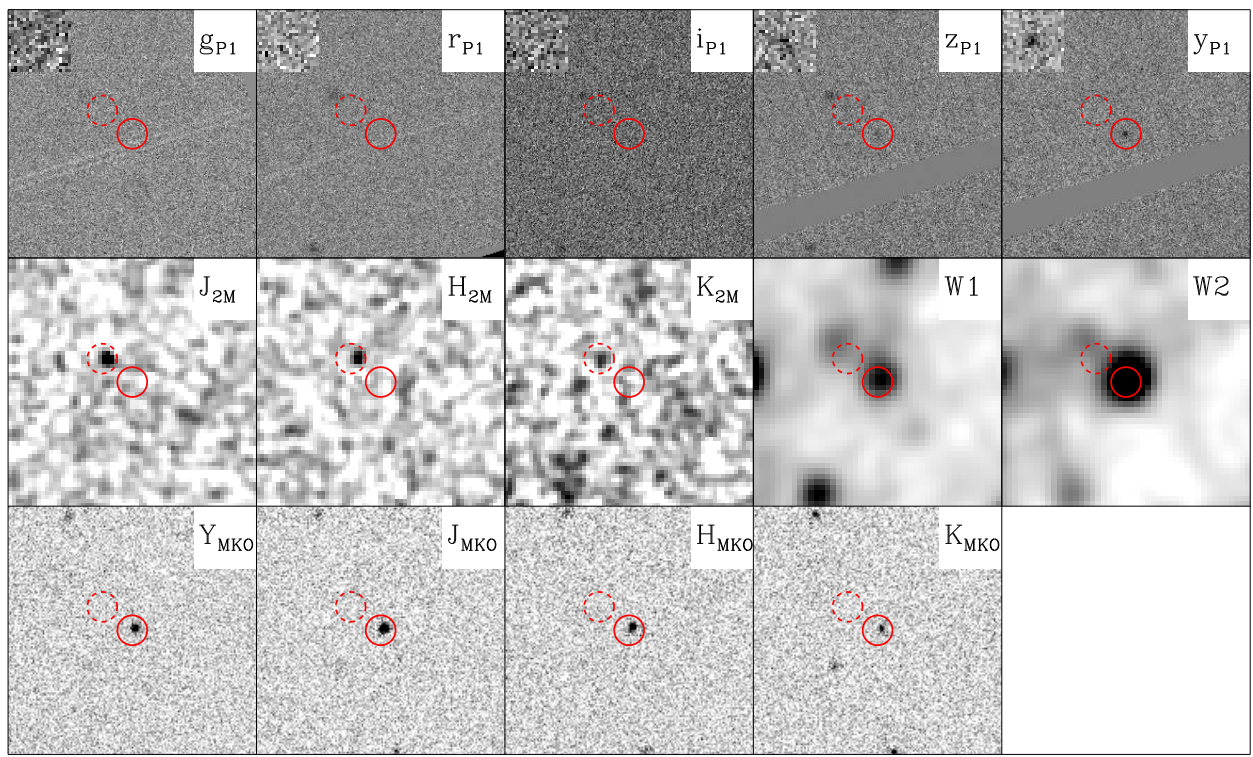}
\caption{\label{LHS2803}Images of LHS~2803B in different filters from Pan-STARRS1 ($g_{P1}$, $r_{P1}$, $i_{P1}$, $z_{P1}$, $y_{P1}$), 2MASS ($J$, $H$, $K_s$), WISE ($W1$, $W2$) and UKIRT ($Y_{MKO}$,$J_{MKO}$,$H_{MKO}$,$K_{MKO}$).The red circles are centered on the Pan-STARRS1 position (center, solid circles, epoch 2011.06) and the 2MASS position (dashed circle, epoch 2001.07). Images are one arcminute across with north up and east left. LHS~2803B was not detected in the $g_{P1}$, $r_{P1}$ and $i_{P1}$ bands. \label{stamps}}

\end{center}
\end{figure}

In order to identify wide common proper motion companions to objects for which we have spectra, we performed a Simbad\footnote{http://simbad.u-strasbg.fr} search within 20 arcminutes of our target coordinates. LHS~2803 \citep{LHS} emerged as an excellent candidate companion to one of our spectroscopically confirmed T dwarfs. Details of the primary and secondary components of this system are listed in Table~\ref{LHS2803system}, and the images of the cool secondary are shown in Figure~\ref{LHS2803}. The separation of the two components is 67.6$\arcsec$ based on their 2MASS positions, the primary is saturated in Pan-STARRS1 and hence we did not attempt to use this source of astrometry or photometry.
\begin{deluxetable}{ccc}
\tablecolumns{3}
\tablewidth{0pc}
\tabletypesize{\footnotesize}
\tablecaption{\label{LHS2803system} The two components of the LHS~2803 system.}
\tablehead{&\colhead{LHS~2803A}&\colhead{LHS~2803B}}
\startdata
Position (J2000)&13 48 07.218 $-$13 44 32.13\tablenotemark{a}\tablenotetext{a}{\cite{Skrutskie2006}}&13 48 02.90 $-$13 44 07.1\tablenotemark{a}\\
Epoch (UT)&2001-01-25\tablenotemark{a}&2001-01-25\tablenotemark{a}\\
$\mu_{\alpha}\cos(\delta)$ ($\arcsec$/yr)&$-$0.675$\pm$0.005\tablenotemark{b}\tablenotetext{b}{\cite{Salim2003}}&$-$0.660$\pm$0.008\tablenotemark{c}\tablenotetext{c}{This work}\\
$\mu_{\delta} ($\arcsec$/yr)$&$-$0.503$\pm$0.005\tablenotemark{b}&$-$0.543$\pm$0.008\tablenotemark{c}\\
$V$ (mag)&15.098$\pm$0.008\tablenotemark{d}\tablenotetext{d}{\cite{Reid2003}.}&\ldots\\
$V-R$ (mag)&1.303$\pm$0.004\tablenotemark{d}&\ldots\\
$R-I$ (mag)&3.044$\pm$0.011\tablenotemark{d}&\ldots\\
$z_{P1}$ (AB mag)&\ldots&21.0$\pm$0.2\tablenotemark{e}\tablenotetext{e}{Zeropoints are calibrated to the PS1 AB magnitude system using the ubercal analysis of \cite{Schlafly2012}, in which photometric data are identified and tied together via overlaps, with nightly zero points allowed to vary.  Additional non-photometric exposures are tied to the ubercal system via relative photometry overlaps (Magnier et al. in prep). The overall system zero points are set to place the photometry on the AB system using photometry of spectrophotometric standards \cite{Tonry2012}.}\\
$y_{P1}$ (AB mag)&\ldots&19.17$\pm$0.12\tablenotemark{e}\\
$J_{2MASS}$ (mag)&10.41$\pm$0.03\tablenotemark{a}&16.48$\pm$0.12\tablenotemark{a}\\
$H_{2MASS}$ (mag)&9.94$\pm$0.02\tablenotemark{a}&16.09$\pm$0.17\tablenotemark{a}\\
$K_{S,2MASS}$ (mag)&9.66$\pm$0.02\tablenotemark{a}&$<$16.45\tablenotemark{a}\\
$Y_{MKO}$ (mag)&\ldots&17.49$\pm$0.05\tablenotemark{c}\\
$J_{MKO}$ (mag)&\ldots&16.39$\pm$0.02\tablenotemark{c}\\
$H_{MKO}$ (mag)&\ldots&16.57$\pm$0.04\tablenotemark{c}\\
$K_{MKO}$ (mag)&\ldots&16.90$\pm$0.08\tablenotemark{c}\\
$W1$ (mag)&9.46$\pm$0.02\tablenotemark{f}\tablenotetext{f}{\cite{Wright2010,WISE2012}}&16.15$\pm$0.07\tablenotemark{f}\\
$W2$ (mag)&9.24$\pm$0.02\tablenotemark{f}&14.18$\pm$0.05\tablenotemark{f}\\
$W3$ (mag)&9.11$\pm$0.03\tablenotemark{f}&$<$12.1\tablenotemark{f}\\
$W4$ (mag)&8.7$\pm$0.3\tablenotemark{f}&$<$9.1\tablenotemark{f}\\
Photometric distance (pc)&21$\pm$3\tablenotemark{d}&$24\pm^{5}_{4}$ pc\tablenotemark{c}\\
$T_{eff}$ (K)&2940$\pm$60\tablenotemark{g}\tablenotetext{g}{\cite{Casagrande2008}}&1100\tablenotemark{h}\tablenotetext{h}{This work, atmospheric model comparison.}\\
&&1120$\pm$80\tablenotemark{i}\tablenotetext{i}{This work, evolutionary model comparison.}\\
Spectral Type&M4.5$\pm$0.5\tablenotemark{c}&T5.5$\pm$0.5\tablenotemark{c}\\
\hline
\enddata
\tablewidth{20pc}
\vspace{-0.7cm}
\normalsize
\end{deluxetable} 
\subsection{Companionship}
\begin{figure}[htbp]
\begin{center}
\epsscale{1.0}
\plotone{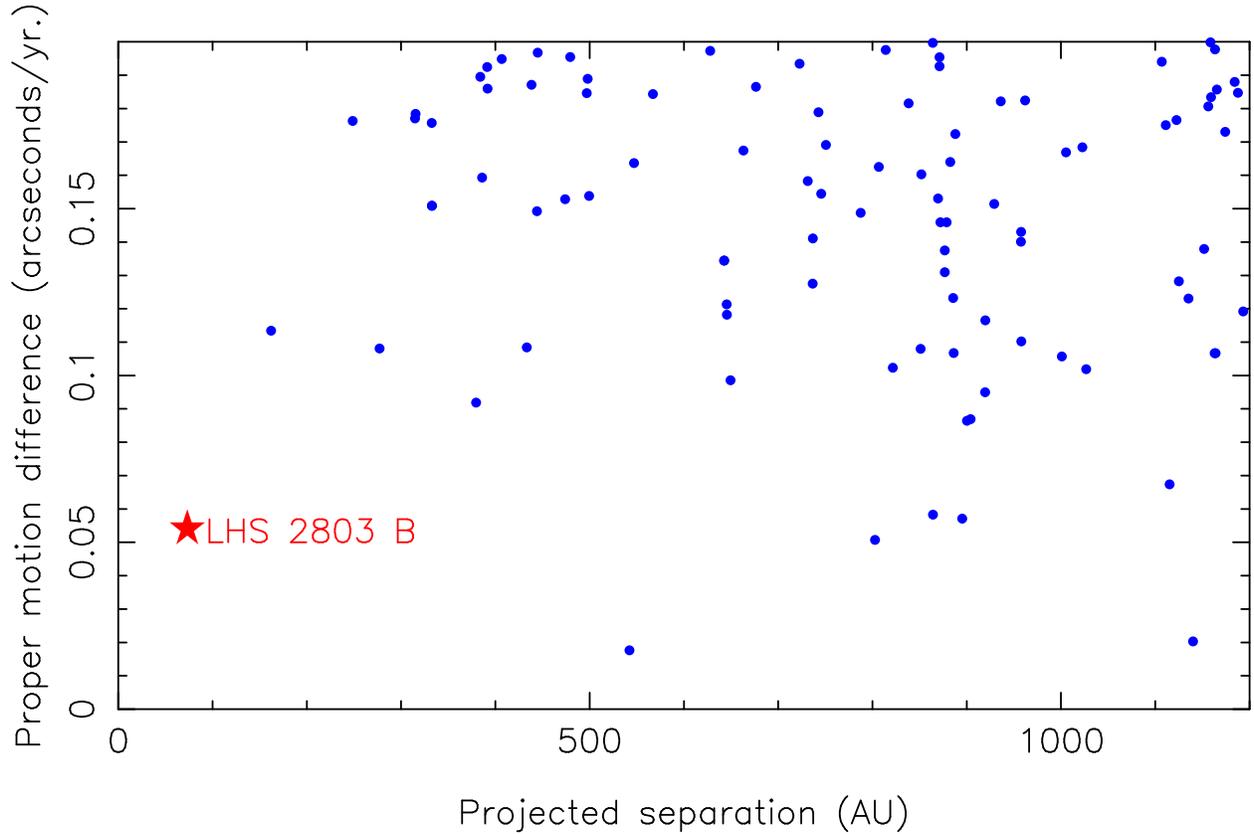}
\caption{\label{LHS2803Bshifted}A comparison of the separation and proper motion difference of the LHS 2803 system (red star) with blue dots showing the chance alignments of objects in our Pan-STARRS1 T dwarf search with stars in the Revised NLTT catalog \citep{Salim2003}. The change alignments are produced using the method of \cite{Lepine2007}. The LHS 2803 system clearly lies in a region of the diagram not occupied by chance alignments.}

\end{center}
\end{figure}

In order to test the likelihood of these two objects being physically associated we applied the criterion described in \cite{Dupuy2012} that the total difference in the proper motion between the two components must be less than one-fifth of the total proper motion of the system. While this is not absolute proof of companionship all known wide ultracool common proper motion companions in the literature meet this criterion. The fractional difference in proper motion for the two components of the LHS~2803 system is 0.05$\pm$0.02,  giving us confidence that the two are likely to be physically associated. In order to further check if these two objects are physically associated, we adopted the approach of \cite{Lepine2007} who also attempt to quantify chance alignments of common proper motion pairs. Their method works by applying an offset to the positions of each star in their catalog and then seeing how many apparent (but false) common proper motion pairs are identified. Their Figure~1 shows the results of this analysis, which extends to proper motions up to 0.45$\arcsec$/yr. As expected, this analysis shows that chance alignments become more common with larger proper motion differences, decreasing total proper motion and wider separations. Using the catalog of objects from our ongoing search for T dwarfs in Pan-STARRS1, we conducted a similar analysis. First to distinguish proper motions from spurious associations between Pan-STARRS1 and 2MASS, we selected only objects with UKIRT follow-up photometry that agreed within 0.6 magnitudes in the $J$ band with their 2MASS $J$ band measurement. This was used by \cite{Deacon2011} as a dividing line between likely true high proper motion objects and spurious associations. We then applied an offset of two degrees in right ascension to each object and paired them with objects in the Revised NLTT catalog of \cite{Salim2003}, yielding a catalog of purely chance associations. The result of this comparison is shown in Figure~\ref{LHS2803Bshifted} along with LHS 2803 B. It is clear that LHS 2803 B lies in the region of the diagram devoid of chance associations. LHS 2803 B also has a significantly higher proper motion than the bulk of the chance associations. Given the probability of chance alignment falls with increasing proper motion, it would seem highly unlikely that this is a coincidence of two unassociated objects.
\section{Follow-up observations}
\subsection{UKIRT/WFCAM}
As part of our follow-up program for candidate T dwarfs identified in Pan-STARRS1, we imaged LHS~2803B on 2011 May 27 (UT) using the WFCAM instrument \citep{Casali2007} on the United Kingdom Infrared Telescope located on Mauna Kea, Hawai`i. Observations were taken in $Y_{MKO}$, $J_{MKO}$, $H_{MKO}$ and $K_{MKO}$ using a five-point dither pattern and five-second integration times per dither in $Y_{MKO}$, $J_{MKO}$ and $H_{MKO}$ and ten-second integrations per dither in $K_{MKO}$. The observations were reduced at the Cambridge Astronomical Survey Unit using the WFCAM reduction pipeline \citep{Irwin2004, Hodgkin2009}. The resulting photometry is shown in Table~\ref{LHS2803system}.
\subsection{NASA IRTF/SpeX}
\label{spectroscopy}
\begin{figure}[htbp]
\begin{center}
\epsscale{1.0}
\plotone{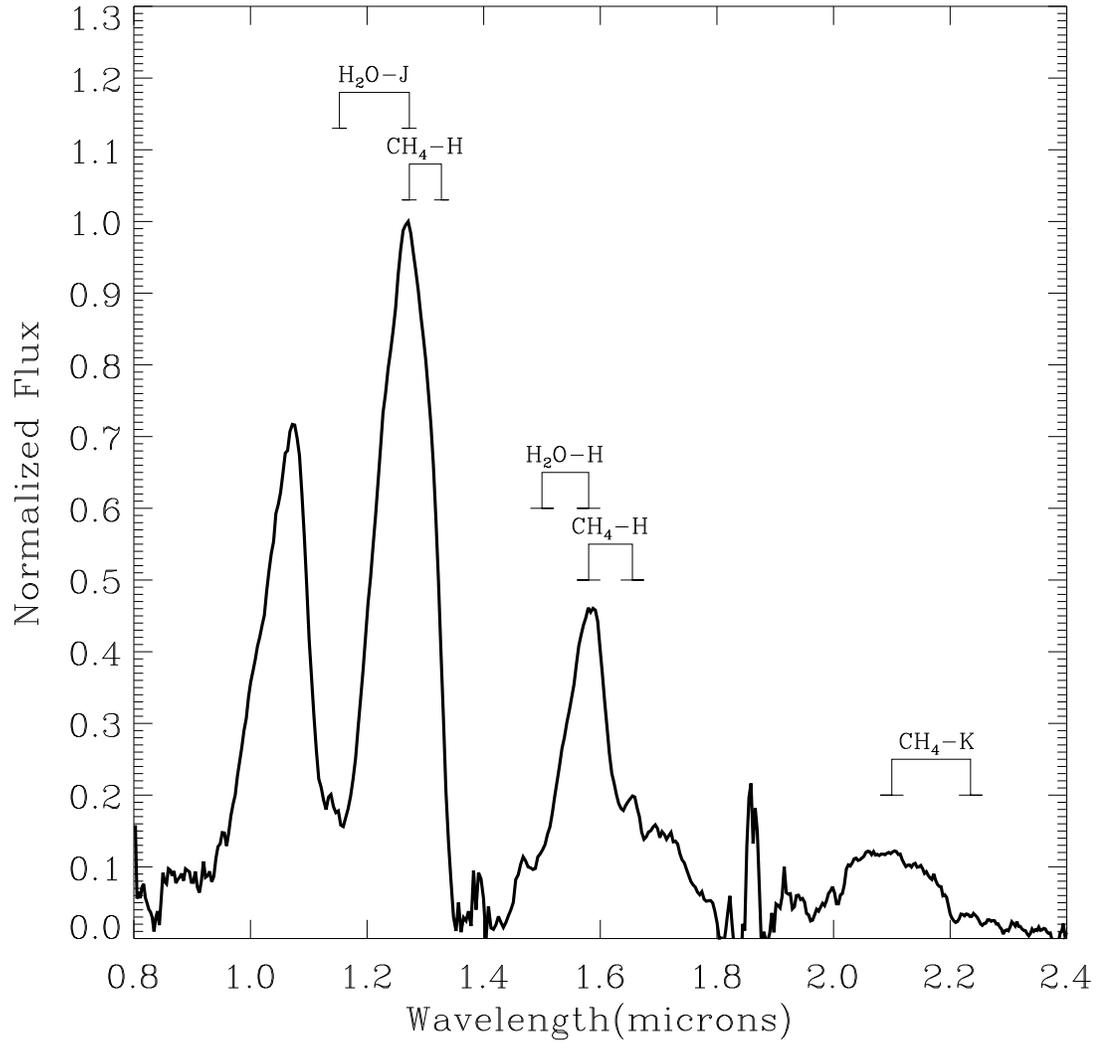}
\caption{\label{LHS2803Bspectrum}The near-infrared prism spectrum of LHS~2803B. The marked molecular bands \citep{Burgasser2006} were used along with visual comparison in the spectral typing of the object. LHS~2803B was assigned a spectral type of T5.5.}

\end{center}
\end{figure}

We obtained spectroscopic observations of LHS~2803~B using the low-resolution prism mode of the SpeX instrument \citep{Rayner2003} on the NASA Infrared Telescope Facility on 2011 May 14 UT. This was done prior to our UKIRT observing run. LHS~2803B was selected for spectroscopic follow-up without near-IR photometry due to its red $z-y$ color, a reliable indicator of an object being a T dwarf \citep{Deacon2011}. Conditions were good during our observations with 0.8$\arcsec$ seeing, hence the 0.8$\times$15 arcsecond slit was used. LHS~2803~B was observed at an airmass of 1.44 with a total integration time of 900s. Individual observations of 90s were taken with the telescope nodded in an ABBA pattern to allow for sky subtraction. The slit was oriented to the parallactic angle to minimize atmospheric dispersion. The data were extracted, telluric corrected and flux calibrated using a contemporaneously observed A0V star and the SpeXTool software package \citep{Cushing2004,Vacca2003}. The resulting spectrum is shown in Figure~\ref{LHS2803Bspectrum} and has a spectral resolution of 82.
\subsection{Spectral Classification}  
LHS~2803B was classified using the flux indices of \cite{Burgasser2006} (see Figure~\ref{LHS2803Bspectrum}) and the polynomial relations of \cite{Burgasser2007}. The individual ratios are given in Table~\ref{spexclass}. This resulted in a spectral type of T5.3$\pm$0.5. Additionally the spectrum was visually compared with IRTF/SpeX prism data for the spectral standards from \cite{Burgasser2006}, resembling an intermediate type between T5 and T6. Hence we assign a final spectral type of T5.5 to LHS~2803B.
\begin{deluxetable}{cccccccc}
\tablecolumns{3}
\tablewidth{0pc}
\tabletypesize{\normalsize}
\tablecaption{\label{spexclass} Near-infrared spectral measurements for LHS2803B.}
\tablehead{\colhead{H2O-J}&\colhead{CH4-J}&\colhead{H2O-H}&\colhead{CH4-H}&\colhead{CH4-K}&\colhead{avg/RMS}&\colhead{Visual}&\colhead{Final}}
\startdata
0.180&0.387&0.279&0.407&0.263&T5.3$\pm$0.5&T5.5&T5.5\\
\hline
\enddata
\normalsize
\end{deluxetable}
\subsection{UH 2.2 m/SNIFS}
Compared to most primary stars of benchmark brown dwarfs, LHS~2803~A is a relatively poorly studied object. The only detailed study of the object comes from \cite{Casagrande2008} who use broadband photometry from \cite{Reid2003} to estimate an effective temperature of 2942$\pm$58 K. 

In order to better classify LHS~2803~A, we obtained an optical spectrum on 2012 April 19 UT in photometric
conditions using the SuperNova Integral Field Spectrograph
\citep[SNIFS;][]{Lantz2004} on the University of Hawaii 2.2-m telescope atop Mauna
Kea. SNIFS is an optical integral field spectrograph with R=1000--1300 that separates the incoming light with a dichroic mirror into blue (3000--5200~\AA) and red (∼5200--9500~\AA)
channels. A single 370~s exposure of the science target at airmass 1.24 was sufficient to achieve
high S/N ($\sim$150) in the red channel. SNIFS data processing was performed with the SNIFS
data reduction pipeline, which is described in detail in \cite{Bacon2001} and
\cite{Aldering2006}. SNIFS processing included dark, bias, and flat-field
corrections, assembling the data into red and blue 3D data cubes, and cleaning them
for cosmic rays and bad pixels. Wavelengths were calibrated with arc lamp exposures
taken at the same telescope pointing as the science data. The calibrated spectrum was
then sky-subtracted, and a 1D spectrum was extracted using a PSF model. Corrections
were applied to the 1D spectrum for instrument response and telluric lines based on
observations of the spectrophotometric standards HZ 21 and Feige 34 \citep{Oke1990}, both taken within 1 hour of the science exposure.

The spectrum was shifted to zero radial velocity by cross-correlating it with
templates from \cite{Bochanski2007}. The spectral type was determined to be M4.5$\pm$0.5 following the
technique of \cite{Lepine2003}, which is based on empirical relations between
spectral type and the spectral indices: TiO5, CaH2, CaH3 \citep{Reid1995}, VO1
\citep{Hawley2002}, and TiO6 \citep{Lepine2003}. This makes LHS~2803A marginally later-type than Wolf 940A \citep[M4, ][]{Reid1995}, the latest previously known star with a wide ($>$100 AU) T dwarf companion \citep{Burningham2009}.
\begin{figure}[htbp]
\begin{center}
\epsscale{1.0}
\plotone{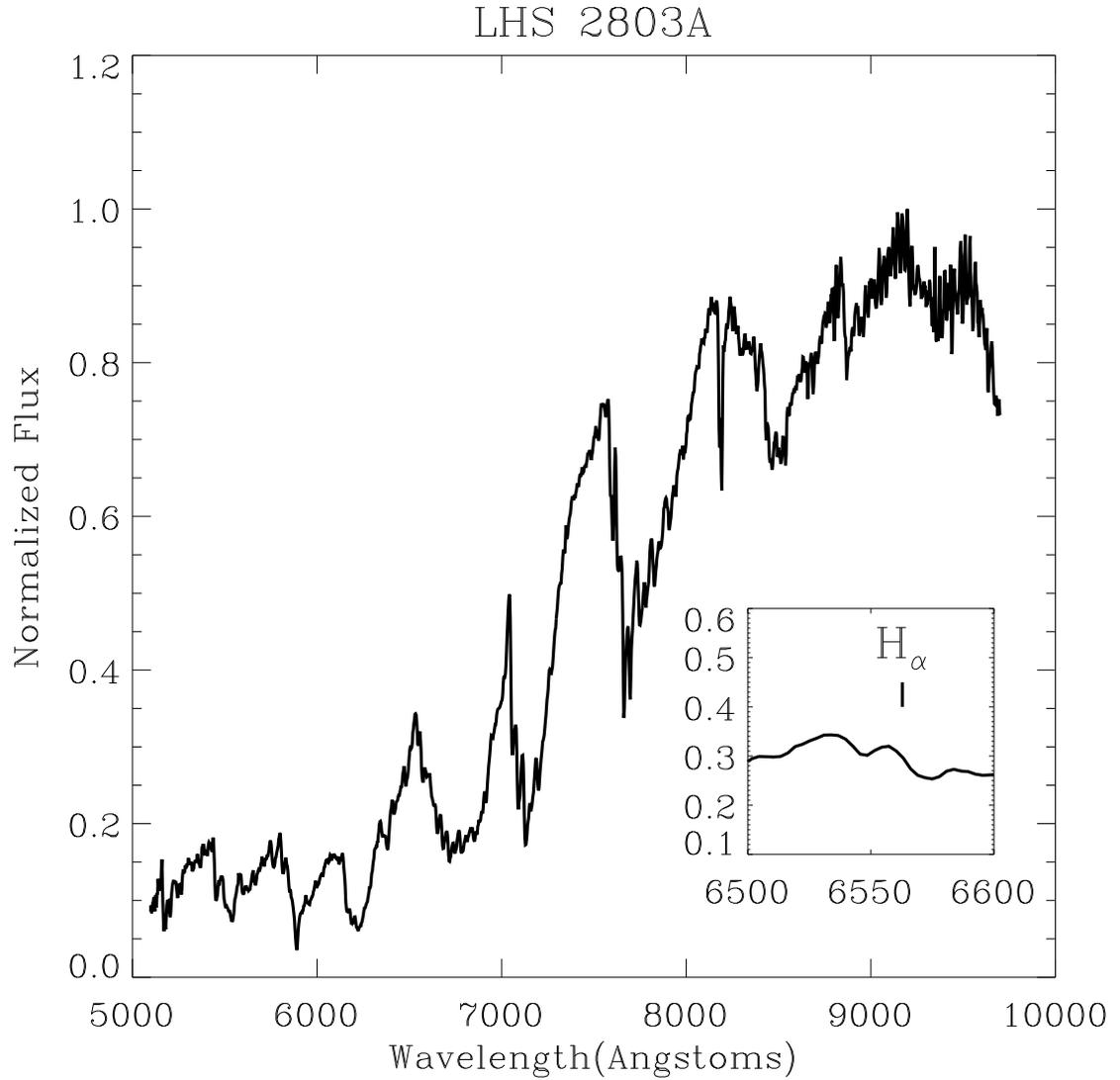}
\caption{\label{LHS2803Aspectrum}Optical spectrum of LHS~2803A. The inset shows the H$\alpha$ region of the spectrum, where there is no strong emission line. The measured equivalent width of the H$\alpha$ is 0.06$\pm$0.33~\AA. Clearly this object cannot be considered to be active. Using the spectral typing scheme of \cite{Lepine2003} we classify this star to be an M4.5 dwarf with an error of 0.5 subclasses.}

\end{center}
\end{figure}
\section{Discussion}

Using our spectral classification of T5.5 together with our UKIRT $Y_{MKO}$, $J_{MKO}$, $H_{MKO}$ and $K_{MKO}$ photometry, we calculated the photometric parallaxes of LHS~2803B using the relations of \cite{Dupuy2012}. These yielded distances of $d_Y=24\pm^{5}_{4}$ pc, $d_J=23\pm^{5}_{4}$ pc, $d_H=23\pm^{5}_{4}$ pc and $d_{K}=26\pm^{4}_{3}$ pc. We adopt the $Y$ band distance (one of the median distances) of $24\pm^{5}_{4}$ pc as our final distance to LHS~2803B.

\subsection{Characterizing LHS~2803~A}

One of the key indicators of youth in M dwarfs is H$\alpha$ emission. We measure the equivalent width of the H$\alpha$ line to be 0.06$\pm$0.33 \AA, indicating that this star is not chromospherically active (see the inset in Figure~\ref{LHS2803Aspectrum}). We compared this result and our derived spectral type of M4.5$\pm$0.5 with the activity lifetimes listed by \cite{West2008}. Hence we take the lower age bound for an M4 star of 3.5 Gyr are being the lower age bound of LHS~2803~A. This lower age bound comes with two caveats. First, the activity lifetimes were calculated using the bulk properties of populations of stars at different Galactic scale heights. It is possible that individual objects may be older/younger than the quoted activity lifetime and still be active/inactive. Second, the activity lifetime - spectral type relation changes rapidly around M5. If LHS~2803~A is on the later end of the uncertainty for our spectral type it would have a substantially older lower age bound (6.5 Gyr for an M5).

We also measured the metallicity index $\zeta$ (originally derived by \citealt{Lepine2007a}, here we use the definition of \citealt{Dhital2012}) to be 1.02. This combined with a TiO5 equivalent width of 0.36 indicated that the object is an M dwarf and not significantly metal poor (see Figure 7 of \citealt{Dhital2012}). 

Unfortunately, LHS~2803 has no trigonometric parallax measurement in the literature. However \cite{Reid2003} calculate a photometric distance of 20.9 pc. They do not quote an error on individual distances; however they estimate their errors to be in the 0.25-0.35 magnitude range. Taking the middle of this range we adopt a photometric distance of 21$\pm$3~pc as the distance to LHS~2803. This distance implies that the system has a separation of 1400$\pm$200 AU and compares well with our calculated photometric distance for LHS~2803B of $24\pm^{5}_{4}$ pc.

Using \cite{Reid2003}'s photometric distance estimate along with the proper motion from \cite{Salim2003} gives a tangential velocity of 83$\pm12$ km/s. Comparing this value with the conditions set out in \cite{Dupuy2012} for kinematic population membership, we determine that LHS~2803A is most likely a member of the thin disk. Thus its age is likely less than 10 Gyr. 

\subsection{Comparison of LHS~2803B with evolutionary models}
\begin{figure}[htbp]
\begin{center}
\includegraphics[scale=.70,angle=90]{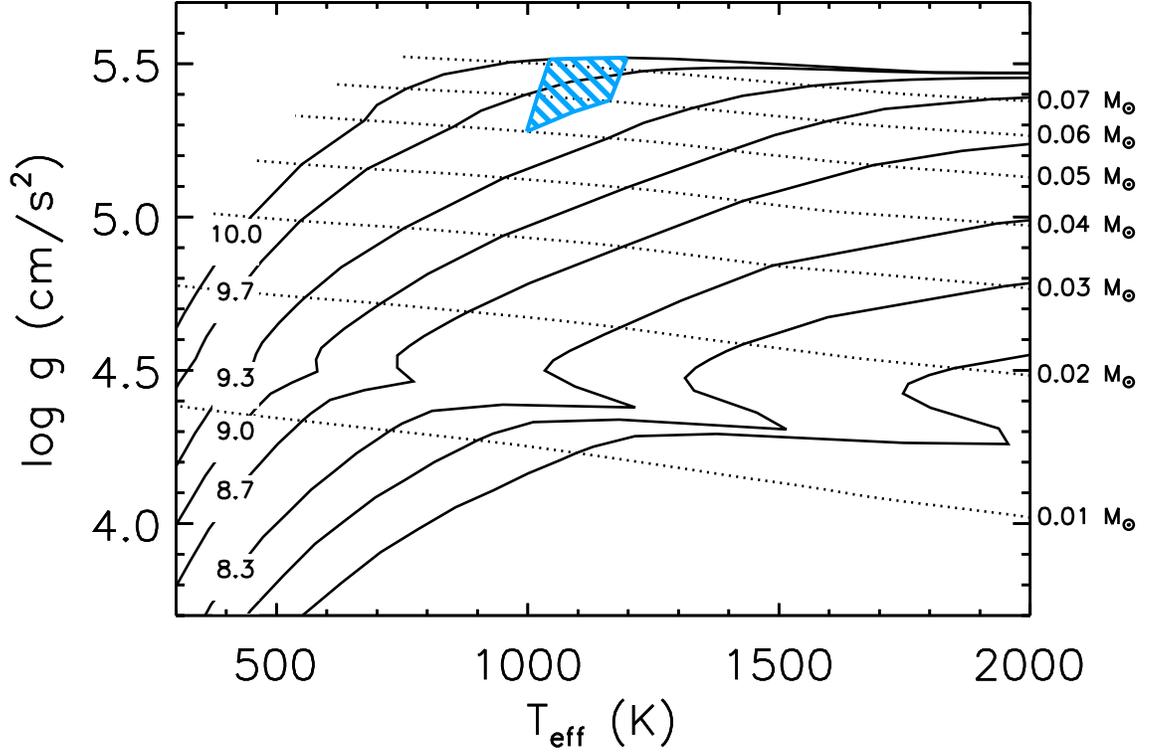}
\caption{Our derived age and bolometric luminosity for LHS~2803B plotted against the evolutionary models of \cite{Burrows1997}. The blue-colored hatched area indicates the constraints on the parameters given the uncertainties on $M_{\rm bol}$ and the age of the primary. Solid lines mark isochrones with the annotated values of $\log$(age). The two unmarked isochrones at the bottom are $\log$(age)=8.0 and 7.7. The dotted lines represent lines of equal mass with the corresponding values noted at their intersection with the righ-hand y-axis. \label{evmodels}}

\end{center}
\end{figure}

Using the \cite{Reid2003} photometric distance, we derived an absolute $H$-band magnitude in the MKO system for LHS~2803B of 15.0 $\pm$ 0.3 mag. Applying the bolometric correction relation from in \cite{Liu2010} gives a bolometric magnitude of 17.4 $\pm$ 0.3 mag. We then compared this value along with our range of ages for the primary (3.5-10.0~Gyr) to the evolutionary models of \cite{Burrows1997}. The resulting constraints on the effective temperature and gravity of LHS~2803B are shown in Figure~\ref{evmodels}.  We also ran Monte Carlo models using our estimated bolometric luminosity and a uniform distribution of ages across our estimated age range. This predicted an effective temperature in the range 1120$\pm$80 K and showed that LHS~2803B appears to be a relatively high mass brown dwarf with $m=72\pm^{4}_{7}$ M$_{Jup}$ and $\log g=5.4\pm0.1$ dex.
\subsection{Comparison of LHS~2803B with atmospheric models} 
We fit the solar metallicity BT-Settl-2010 models \citep{Allard2010} to our SpeX/prism spectrum of LHS~2803B following the $\chi^2$ minimization
technique described in \cite{Cushing2008} and \cite{Bowler2009}. The spectrum was first flux calibrated to the UKIDSS $J$-band photometry, which is the most precise of the UKIDSS photometric measurements for LHS~2803B. We use the 0.8--2.4~$\mu$m region in the fits, but exclude the 1.60-1.65~$\mu$m window because of incomplete methane line lists in the models \citep{Saumon2007} and also the 1.80--1.95~$\mu$m region, which suffers from strong telluric absorption. The grid of 42 models spans effective temperatures from 500--1500~K with 100~K intervals and $\log g$ from 4.0--5.5 dex in steps of 0.5 dex. The prism spectrum contains 387 data points, which dominates the selection of the best-fitting model, over the small number of photometric measurements (like PS1 or WISE). Hence we did not use the photometric data points in our model spectrum selection process.

\begin{figure}[htbp]
\begin{center}
\includegraphics[scale=.70]{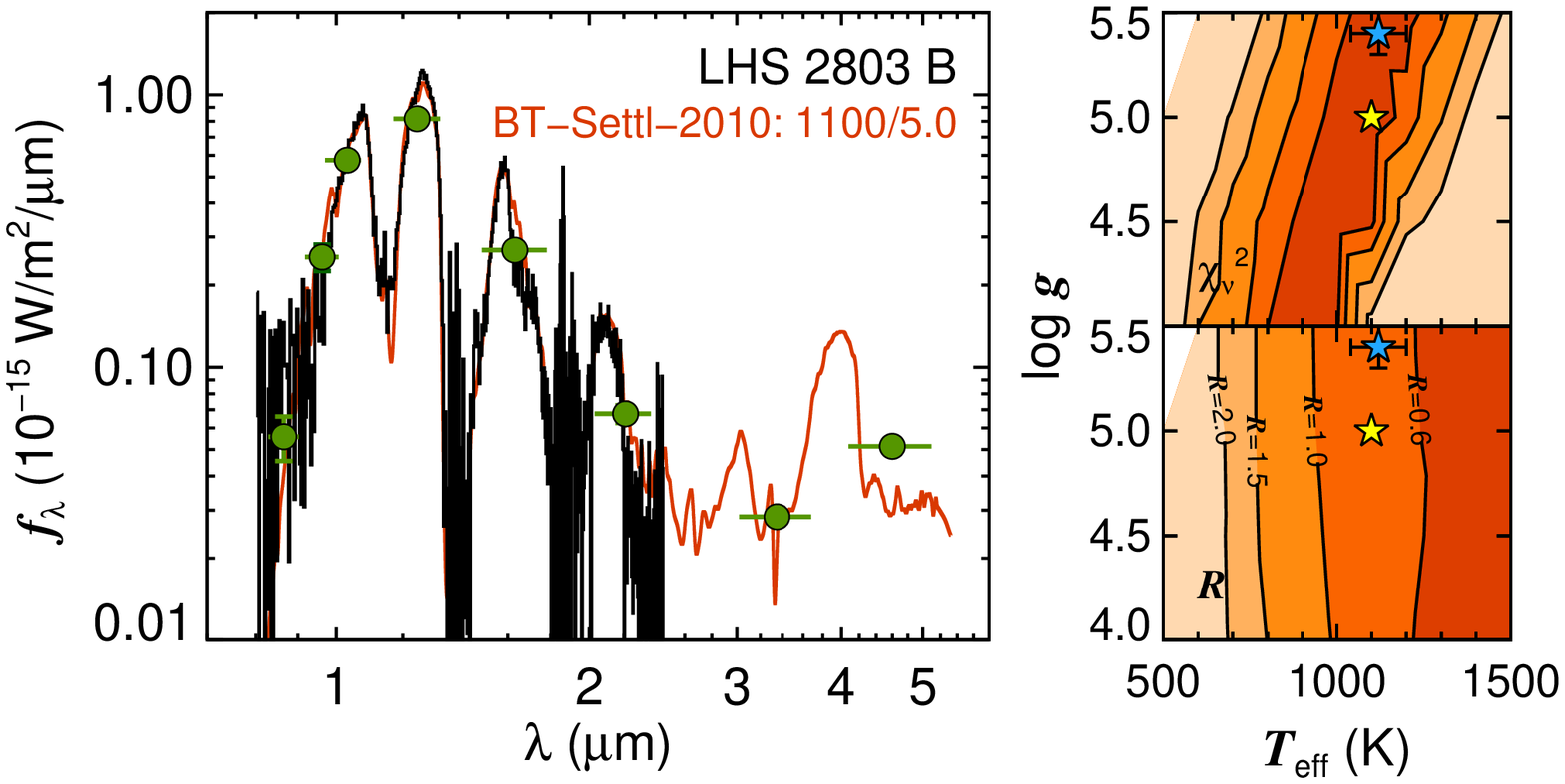}
\caption{Comparison of the best-fit BT-Settl-2010 model spectrum
(red) to our SpeX/prism spectrum of LHS~2803B (black) and 0.8--5~$\mu$m photometry (green). The $z$ and $y$-band photometry are from PS1, the $YJHK$ photometry are from our UKIRT observations, and the 3.3 and 4.6~$\mu$m photometry are from WISE \citep{Wright2010}. Zero point flux denities for PS1, UKIRT, and {\it WISE} are from \cite{Tonry2012}, \cite{Hewett2006}, and \cite{Wright2010}, respectively.  The photometric uncertainties are smaller than the size of the symbols. \textit{Right}: Reduced $\chi^2$ surface of the model atmosphere fitting (top) and the inferred radius (in $R_\mathrm{Jup}$) from the fits (bottom; see \citealt{Bowler2009} for details).  Yellow stars show the best-fitting model atmosphere parameters and blue stars show the results from evolutionary models.  The best-fit parameters are within one grid step of the evolutionary model predictions. \label{atmodels}}

\end{center}
\end{figure}

The best-fit model atmosphere has $T_\mathrm{eff}$=1100~K and log~$g$=5.0.  Given the relatively coarse grid spacing of the models, this result agrees quite well with the predictions from evolutionary models ($T_\mathrm{eff}$=1120~K~$\pm$80~K, log~$g$=5.4~$\pm$~0.1). Fits to the T4.5 companion HD~38939B in \cite{Deacon2012} using the  same fitting technique and model atmospheres also resulted in best-fit parameters that were within one grid step of the evolutionary model predictions. Figure~\ref{atmodels} shows the best-fit model compared to the prism spectrum and the  0.8--5~$\micron$ photometry.  The model is an excellent match to the near-infrared data, but disagrees with the WISE $W2$-band photometry.
\section{Conclusions}
We have identified a common proper motion companion to the M dwarf LHS~2803. Spectroscopic observations classify the primary as M4.5$\pm$0.5 and the secondary as T5.5$\pm$0.5. Based on the literature photometric distance of 21$\pm$3~pc, we calculate a projected separation of 1400$\pm$200 AU. This system is among the widest known substellar companions to an M dwarf. We use the primary's lack of H$\alpha$ emission to set a lower limit of 3.5 Gyr on the system age and its disk kinematics and approximately solar metallicity to set an upper age bound of 10 Gyr. Based on this age range, evolutionary model calculations indicate that the secondary has $T_\mathrm{eff}$=1120~$\pm$80~K and log~$g$=5.4~$\pm$~0.1 with a mass of $72\pm^{4}_{7}$~M$_{Jup}$, suggesting a relatively old, higher gravity object close to the maximum possible mass for a T dwarf. Comparing atmospheric models to our near-IR spectrum give a best fit of 1100~K and log~$g$=5.0, within one grid point of our evolutionary model calculations. The effective temperatures from the evolutionary and atmospheric models are in good agreement, in common with the similar benchmark mid-T dwarfs HN~Peg~B and HIP~38939~B \citep{Deacon2012}. Pan-STARRS1 is ideally suited for identifying substellar companions to nearby stars, and a dedicated effort is underway to identify such objects.
\acknowledgments
The PS1 Surveys have been made possible through contributions of the Institute for Astronomy, the University of Hawaii, the Pan-STARRS Project Office, the Max-Planck Society and its participating institutes, the Max Planck Institute for Astronomy, Heidelberg and the Max Planck Institute for Extraterrestrial Physics, Garching, The Johns Hopkins University, the University of Durham, the University of Edinburgh, Queen's University Belfast, the Harvard-Smithsonian Center for Astrophysics, and the Los Cumbres Observatory Global Telescope Network, Incorporated, the National Central University of Taiwan, and the National Aeronautics and Space Administration under Grant No. NNX08AR22G issued through the Planetary Science Division of the NASA Science Mission Directorate. They would also like to thank Dave Griep for assisting with the IRTF observations. This research has benefited from the SpeX Prism Spectral Libraries, maintained by Adam Burgasser at http://www.browndwarfs.org/spexprism. This publication makes use of data products from the Two Micron All Sky Survey, which is a joint project of the University of Massachusetts and the Infrared Processing and Analysis Center/California Institute of Technology, funded by the National Aeronautics and Space Administration and the National Science Foundation. This research has benefited from the M, L, and T dwarf compendium housed at DwarfArchives.org and maintained by Chris Gelino, Davy Kirkpatrick, and Adam Burgasser. E.A.M. and M.L. were supported by NSF grant AST 0709460. E.A.M. was
also supported by AFRL Cooperative Agreement FA9451-06-2-0338. This project was supported by DFG-Sonderforschungsbereich 881 "The Milky Way System". This publication makes use of data products from the Wide-field Infrared Survey Explorer, which is a joint project of the University of California, Los Angeles, and the Jet Propulsion Laboratory/California Institute of Technology, funded by the National Aeronautics and Space Administration.The United Kingdom Infrared Telescope is operated by the Joint Astronomy Centre on behalf of the Science and Technology Facilities Council of the U.K. 
This paper makes use of observations processed by the Cambridge Astronomy
Survey Unit (CASU) at the Institute of Astronomy, University of Cambridge.
 The authors would like to thank Mike Irwin and the team at CASU for making the reduced WFCAM data available promptly and Tim Carroll and Watson Varricatt for assisting with UKIRT observations. This research has made use of the SIMBAD database,
operated at CDS, Strasbourg, France.\\
{\it Facilities:} \facility{IRTF (SpeX)}, \facility{PS1}, \facility{UKIRT (WFCAM)}, \facility{UH:2.2m (SNIFS)}, 
\bibliography{ndeacon}
\bibliographystyle{apj}
\end{document}